# Learning-based Link Prediction Methods Integrating Network Topological Features and Embedding Representations

JIN Zi-Xuan[1], YI Jun-Fan[1], SHANG Ke-Ke[1*]

[1](Computational Communication Collaboratory, Nanjing University, Nanjing 210023, China)

**Abstract:** Link prediction, as a frontier task in complex network topology analysis, aims to infer the existence of latent links between node pairs based on observed nodes and structural information. We propose an ensemble link prediction model that integrates network topology features and embedding representations (TELP), designed to overcome the limitations of conventional heuristic methods in capturing node attributes and deep structural patterns, as well as the weak interpretability and limited generalization of learning-based approaches. TELP leverages a multi-stage architecture. Local connectivity patterns are captured through network-type-aware selection of homogeneous and heterogeneous topology features, which also promotes interpretability. To incorporate global structure, Node2Vec embeddings are generated and fused with these topology features, resulting in comprehensive multi-dimensional representations. Building on this enriched feature space, an ensemble of logistic regression, random forest, and XGBoost models is deployed to maximize predictive performance and robustness. Experiments on nine classical benchmark networks demonstrate that TELP achieves superior AUC and AP performance compared with traditional heuristic approaches and mainstream graph neural network models, while ablation studies further confirm that feature fusion and ensemble strategies are essential for optimal performance.

# 1 Introduction

In complex networks, link prediction aims to estimate the likelihood of a link between two nodes based on observed node information and topological structure[1]. This task includes the recovery of missing or fake links in static networks and the prediction of future or vanishing connections in dynamically evolving networks[1]. The problem has further expanded to modeling link formation and dissolution driven by node addition or removal in temporal and growing networks[2]. Originating in the field of computer science[1], link prediction has attracted widespread attention across disciplines. In biological research, it is used to uncover unknown protein–protein interactions and latent patterns in metabolic networks[3]; in social network analysis, it helps identify potential social relationships and anticipate information diffusion paths[4]; and in engineering applications, it supports tasks such as recommender systems[5] and knowledge graph completion[6].

Early link prediction studies primarily relied on heuristic algorithms, which compute node similarity by exploiting statistical and physical characteristics embedded in network topologies. Node similarity in networks can be characterized through node attributes or structural similarity [7], while classical heuristic methods largely depend on the latter. Representative similarity indices, such as Common Neighbors[1] are mostly grounded in the theory of triadic closure[7]. However, as research progressed, scholars observed that many real-world networks are heterogeneous with low clustering coefficients, where triadic-closure-based approaches often perform poorly or fail. To address these limitations, new indices such as the Heterogeneity Index (HI) were proposed[8]. Although heuristic algorithms are effective in capturing and interpreting structural patterns, they suffer from limited generalizability, an inability to incorporate node attributes, and inadequate capacity to learn deeper structural representations. Consequently, they are increasingly insufficient for meeting the demands of link prediction in modern complex networks.

To overcome the inherent limitations of heuristic link prediction algorithms, learning-based approaches emerged. Early studies relied on probabilistic models and maximum likelihood estimation to characterize the statistical regularities underlying potential links[9]. Later, Hasan et al.[10]introduced a machine-learning perspective by formulating link prediction as a binary classification problem. Building on this idea, subsequent research incorporated hand-crafted topological features within various theoretical models. Representative examples include the inductive matrix completion mechanism proposed by Zhao et al.[11]and the high-performance supervised model developed by Lichtenwalter et al.[12], which integrates multiple structural features for enhanced predictive accuracy.However, traditional machine-learning-based link prediction methods still face notable challenges. First, these approaches rely heavily on manual feature engineering, requiring domain knowledge and heuristic rules to construct effective features, which makes it difficult to capture latent high-order structural patterns in complex networks[13]. Second, their generalization performance often depends on the distribution of the training data and the design of specific heuristic rules, resulting in degraded performance on networks with strong heterogeneity or substantial structural differences[14].

In recent years, the rapid development of deep learning for graph-structured data has significantly advanced learning-based link prediction. Algorithms built upon graph neural networks (GNNs)[15] and their variants, such as GCN (Graph Convolutional Network)[16], GAT (Graph Attention Network)[17], and GraphSAGE (Sample and Aggregate)[18], enable end-to-end feature propagation and aggregation, allowing models to automatically extract multi-level structural patterns and node attribute signals during training. These methods can therefore capture topological information from local neighborhoods to global structures while eliminating the reliance on manual feature engineering, leading to strong performance on large-scale complex networks.

Despite these advantages, GNN-based link prediction methods inherit several challenges associated with deep learning models. First, their multi-hop aggregation mechanisms rely on complex nonlinear transformations, making it difficult to establish clear correspondences between learned representations and known physical or structural properties of networks. This results in limited model interpretability. Second, their performance is highly sensitive to architectural choices and hyperparameters, such as the number of aggregation layers, neighbor sampling strategies, and regularization strength, which often constrains generalization across networks with varying structural characteristics. Third, training these models typically involves large-scale matrix operations and graph convolutions, imposing substantial computational and memory requirements, and thus making them heavily dependent on high-performance hardware resources[19].More importantly, these limitations restrict the transferability and applicability of deep-learning-based link prediction in domains such as the social sciences, where link prediction is widely used. In such contexts, researchers value not only predictive accuracy but also theoretical interpretability and the ability to uncover underlying behavioral mechanisms, ensuring that model outputs align with social processes and domain insights[20]. Furthermore, the demand for methodological transparency and operational convenience makes resource-intensive GNN models difficult to integrate into mainstream analytical workflows in these fields.

In this context, graph-embedding-based approaches have regained attention as a more transparent and flexible alternative. Network embedding methods[21] map high-dimensional and highly irregular topological structures into a low-dimensional continuous vector space. Compared with traditional heuristic indices, these representations not only preserve local neighborhood patterns but are also capable of capturing higher-order structural properties, such as community organization, path dependencies, and role similarities, that are essential for accurate link inference. Consequently, embedding-based predictors often achieve notably stronger performance while maintaining a higher degree of interpretability. Moreover, the vectorized representation enables more flexible feature integration in downstream models, allowing a balance between expressive power and computational efficiency.

Building on these advantages, we proposes an integrated link-prediction model that leverages structural heterogeneity in a principled manner. Specifically, we develop an ensemble model, named TELP (Topology- and Embedding-based Link Prediction), that combines handcrafted topological features with learned embedding representations. TELP is designed to retain the interpretability characteristic of heuristic measures while exploiting the high-order structural expressiveness offered by embedding methods. As a result, TELP aims to deliver robust predictive performance, improved generalizability, and enhanced explanatory value across diverse network environments.

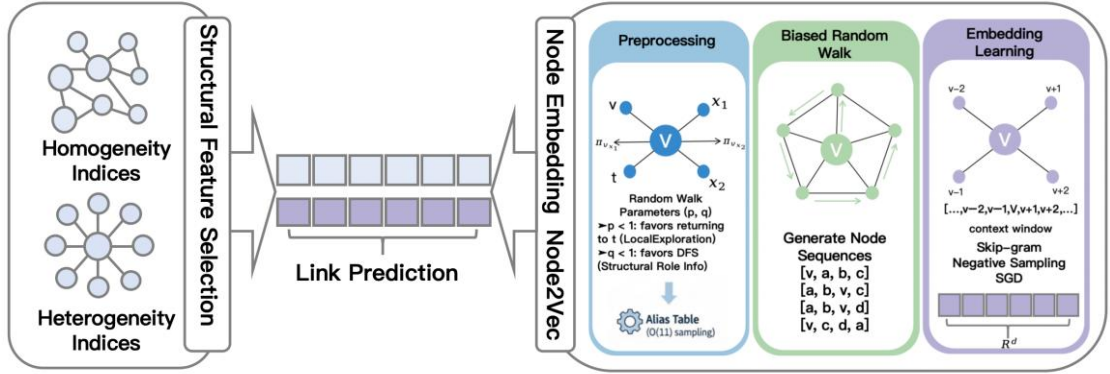

Figure 1. Overall architecture of the TELP framework. *The framework consists of two core feature-construction modules. The first is the structural feature selection module, in which the network is categorized into homogeneous and heterogeneous types, and local structural information is extracted based on homogeneity indices and heterogeneity indices, respectively. The second module is the Node2Vec-based node embedding component, which includes preprocessing, biased random walks, and Skip-gram training with negative sampling to obtain high-order structural representations. Finally, these two categories of features are combined to construct feature vectors for candidate node pairs, enabling the link prediction task.*

As illustrated in Fig. 1, the feature-selection stage begins with a structural categorization of the target network into homogeneous and heterogeneous types. For homogeneous networks, the local topological structures are characterized using four classical similarity-based indices: Common Neighbors (CN)[1], Jaccard coefficient [22], Adamic–Adar index[23], and Resource Allocation (RA) index[24]. In contrast, for heterogeneous networks, we incorporate the Heterogeneity Index (HI)[8] to capture the asymmetric linking tendencies induced by hub-like nodes.

On this basis, Node2Vec[25] is applied to each training subgraph to generate low-dimensional node embeddings, which are subsequently combined using four standard vector-operator schemes to construct pairwise representations. These embedding-based features are then concatenated with the selected topological metrics to form the final feature vector for each candidate link.

To balance interpretability with nonlinear modeling capacity, we adopt Logistic Regression[26], Random Forest[27], and XGBoost[28] as base learners, and further integrate them through an ensemble-learning mechanism. This design enhances predictive accuracy while also improving model stability and generalizability across network types. The main contributions of this work can be summarized as follows:

1. We propose a topology and structure-aware link prediction model that adaptively selects appropriate structural indices based on network homogeneity, and integrates them with embedding-based features to strengthen cross-network applicability.

2. We develop a modular pairwise-representation strategy that systematically combines Node2Vec embeddings with multiple vector operators, and empirically demonstrate the irreplaceable role of embedding features in capturing latent structural regularities.

3. We design an ensemble architecture combining linear and tree-based models, achieving a favorable trade-off between interpretability and expressive power, and yielding more stable predictions under complex feature interactions.

4. We conduct extensive experiments on nine widely-used real-world datasets from complex network analysis and computer science. Comparisons against five heuristic baselines and three classical neural network models show that the TELP consistently and substantially outperforms existing methods in predictive accuracy and robustness.

## 2 Related work

### 2.1 Heuristics-based link prediction

Early studies on link prediction predominantly relied on heuristic algorithms that operate directly on network topology without complex training procedures. These methods estimate the likelihood of a missing or future link by computing similarity scores between node pairs, which are generally categorized into three classes: local similarity indices, path-based measures, and random-walk-based metrics[7]. Among these, local similarity indices have been the most widely used due to their simplicity and grounding in the principle of triadic closure, which assumes that two nodes sharing many common neighbors are more likely to form a link. The Common Neighbors (CN) index, introduced by Liben-Nowell and Kleinberg[1], represents the most fundamental form of this class. Subsequent studies incorporated node-degree information and proposed various extensions, including Salton's index[29], the Jaccard coefficient[22], and the Sorenson index[30], thereby enriching the range of local topological descriptors. Further enhancements were proposed by Zhou, Lü and colleagues, who developed the Resource Allocation (RA) index[24], which assigns greater weight to low-degree intermediate nodes, and by Lü et al.[31], who extended heuristic methods to weighted networks through the incorporation of link strength and weak-tie theory.

However, as research progressed, it became evident that many real-world networks exhibit strong heterogeneity, which is manifested by multi-type nodes and edges or highly skewed degree distributions. This condition often undermines the effectiveness of closure-based heuristics. To address this issue, Shang et al. emphasized the importance of degree heterogeneity in sparse complex networks and proposed integrating heterogeneity-oriented indices to better capture hub-driven connection patterns[8].

Heuristic methods are advantageous for their interpretability, computational efficiency, and scalability to large networks. However, their limitations are equally prominent. First, their reliance solely on topological information prevents the incorporation of node attributes. Second, most heuristics capture only linear or shallow structural patterns, making them insufficient for networks with complex, high-order dependencies[9]. Consequently, while heuristic approaches laid the foundational groundwork for link prediction research, their predictive power declines significantly when confronted with the increasing structural complexity of modern networks.

**2.2 Learning-based link prediction**

To address the limitations of heuristic link prediction methods, researchers increasingly adopted learning-based approaches. In this vein, methods based on probability and maximum likelihood served as a key transitional development in the shift from heuristic to learning-based algorithms. Probabilistic Relational Models (PRMs) constitute a major class, encompassing models such as Relational Bayesian Networks (RBNs)[32], Relational Markov Networks (RMNs)[33], and Relational Dependency Networks (RDNs)[34], along with the Directed Acyclic Probabilistic Entity-Relationship (DAPER) models. Among maximum-likelihood estimation (MLE) algorithms, the Hierarchical Random Graph (HRG) model[35] and the Stochastic Block Model (SBM)[36] are the most representative, both of which estimate link probabilities between node pairs by maximizing the likelihood function. These approaches are capable of revealing network structure properties such as community partitions, hierarchical organization, or group attributes, thus providing deeper structural interpretations alongside the prediction of missing links. However, their inference procedures often involve complex parameter estimation and probabilistic optimization, resulting in high computational complexity that limits their direct applicability to large-scale networks.

With advances in machine learning methodology, link prediction was subsequently formalized as a supervised learning task. In this model, potential node pairs are treated as training samples, represented by feature vectors that characterize their relationship, and a classifier is employed to learn the discriminative boundary between connected and unconnected pairs[37]. Feature selection and construction are central to this approach. The mainstream practice involves extracting general metrics from the network topology, including local-neighborhood indices such as Common Neighbors (CN) and Adamic–Adar (AA), path-based features like the Local Path index (LP) and Katz index, as well as degree-related statistics[7].

Building on this foundation, researchers have also attempted to incorporate node attributes and edge semantic information into feature construction to overcome the limitations of relying solely on topological features. For example, Hasan[10] utilized the degree of keyword overlap as a supplementary feature in co-authorship networks, demonstrating a strong correlation between semantic similarity and collaborative relationships. Similar studies include the integration of user interest tags in social networks[38] and functional annotations in biological interaction networks[39]. While these features often yield effective improvements in predictive performance, their inherent reliance on domain-specific knowledge compromises their generalizability.

A wide array of classifiers has been applied in this setting [40], including logistic regression, support vector machines (SVMs), decision trees, random forests, and Naïve Bayes. Compared with individual heuristic methods, the supervised learning model can integrate multi-source information and automatically adjust feature weights during training, thus generally achieving superior accuracy and robustness. However, this paradigm still suffers from intrinsic bottlenecks: Firstly, feature engineering relies heavily on researchers' prior knowledge and manual design, a constraint that may overlook complex, latent patterns in the network and thus limit prediction accuracy[14]; Secondly, feature construction can face prohibitive computational complexity in large-scale networks—issues that ultimately motivated the development of embedding-based approaches.

Network embedding methods subsequently emerged as a powerful alternative within the learning-based paradigm. By learning low-dimensional vector representations of nodes in an automated manner, they eliminate the need for laborious manual feature design and overcome its inherent limitations.

Early embedding algorithms, exemplified by DeepWalk[41], treated node sequences generated by random walks analogously to sentences in natural language and utilized the Skip-gram model to train dense vector representations for nodes, thereby encoding higher-order proximity. Node2Vec [42]further enhanced random-walk sampling by introducing biased depth-first (DFS) and breadth-first (BFS) strategies. This improvement enables the model to flexibly balance between local and global structural information, and experimental results demonstrated its superior link prediction performance relative to DeepWalk. In addition to random-walk-based methods, matrix-factorization approaches, such as GraRep[43] and HOPE[44], directly decompose adjacency or similarity matrices to acquire vectorized representations of multi-order proximity. More recent advances include complex-valued embedding models. For instance, ComplEx[45], proposed by Trouillo et al., utilizes complex vector decomposition of tensors, allowing it to simultaneously model both symmetric and antisymmetric connectivity patterns.

Overall, embedding methods effectively preserve the topological proximity of the network within a continuous latent space, significantly enhancing the expressive capacity and generalization ability of learning-based link prediction. Furthermore, they established the theoretical and methodological foundation for subsequent deep learning and Graph Neural Network approaches.

In recent years, Graph Neural Networks (GNNs) have become a dominant paradigm for link prediction, driven by the rapid progress of deep learning in graph representation learning. By leveraging a message-passing mechanism, GNNs [15] automatically learn node representations and high-order structural information by propagating and aggregating features between nodes and their neighbors via a message-passing mechanism.

Early efforts in this direction include the Graph Convolutional Network (GCN)[16], which aggregates features from neighboring nodes using either spectral domain convolution or approximated spatial convolution, thereby effectively capturing both local and global topological relationships. Subsequently, the Graph Attention Network (GAT)[17] introduced an attention mechanism during the feature aggregation stage. This modification allows the model to adaptively assign weights based on the importance of neighboring nodes, thereby enhancing the flexibility and expressive capacity of information aggregation. GraphSAGE[18], through the combination of neighbor sampling and various aggregation functions, achieved inductive learning on large-scale networks, which provides improved scalability for link prediction in dynamic and massive networks. Building upon these foundational models, subsequent researchers have proposed various extensions to accommodate more complex network structures and diverse task scenarios, thereby further expanding the application scope of GNNs in link prediction.

Nevertheless, despite their strong representational power, GNN-based link predictors face several well-documented limitations inherent to deep learning models. First, the process by which these algorithms aggregate multi-hop neighborhood information largely relies on black-box, non-linear combinations, which lack an explicit connection to underlying network topological properties, leading to insufficient interpretability. Second, GNNs are prone to performance degradation in scenarios marked by scarce node attributes or strong network heterogeneity. Furthermore, their sensitivity to model architecture and hyperparameter choices results in generalization difficulties. Third, the reliance of model training on extensive computational resources further leads to sub-optimal application transfer performance.

Compared to heuristic methods, learning-based approaches offer the distinct advantage of integrating multi-dimensional features, thereby overcoming the constraint of relying solely on topology. They further demonstrate significant potential in modeling non-linear relationships and facilitating cross-domain applications. However, these methods concurrently grapple with inherent drawbacks, including sub-optimal interpretability, restricted generalization capability influenced by hyperparameter settings, and a significant reliance on computational resources for model training. Consequently, the integration of the respective strengths of heuristic and learning-based prediction algorithms, aimed at constructing a simple yet highly efficient link prediction model that balances interpretability with expressive power, represents a crucial current research trend and a potential area for breakthrough. To this end, we proposes an algorithm that differentially selects between network structural features and learned node embeddings based on the specific type of network topological structure. By utilizing an ensemble learning approach to enhance predictive robustness, our model aims to simultaneously strengthen the model's expressive capacity and interpretability.

## 3 Methodology

### 3.1 Problem Statement

We define an undirected static graph $G=(V, E, A, X)$ with N nodes. Here, $V=v_1, v_2, v_3, \ldots v_N$ denotes the set of nodes; $E=\{e_{ij}|v_i, v_j \in V\}$ is the set of observed edges, indicating whether nodes $v_i$ and $v_j$ are connected. The adjacency matrix $A \in \{0, 1\}^{N*N}$ is defined such that:

$$A_{ij} = \begin{cases} 1, & \text{if } v_i \text{s connected to } v_j, \\ 0, & \text{otherwise.} \end{cases}$$

The degree matrix $D = \{d_1, d_2, \ldots, d_N\}$ specifies the degree of each node. In addition, for every potential edge $(v_i, v_j)$, our method constructs a multi-dimensional feature vector that integrates structural features and node embedding features. In this network, no multiple edges or self-loops exist, and some links may be unobserved or missing. The goal of link prediction is to estimate the probability that these unobserved node pairs will form connections based on the fixed structural information of the graph.

Typically, to evaluate algorithm performance, a certain percentage of the observed edges is selected as the test set $E^P$, while the remaining edges and all nodes form the training set $E^T$. This training set is utilized for calculating similarity scores between node pairs, constructing feature vectors, and learning model parameters. We ensure that $E = E^P \cup E^T$ and $E^P \cap E^T = \emptyset$。 Within the complex networks domain, the standard practice is to assign 10% of the edges to the test set and 90% to the training set. In the field of computer science, five-fold cross-validation[46] is a widely applied data validation procedure. This technique randomly and uniformly partitions the original dataset into five subsets of approximately equal size. In each cycle, one subset is designated as the test set for final assessment of the model's generalization capability, while the remaining four subsets are merged to form the training set for parameter learning and model fitting. This process is repeated five times, ensuring that each subset serves as the test set exactly once. This procedure fully assesses the model's generalization ability across multiple divisions, mitigating the potential randomness inherent in a single split. To ensure the stability and validity of the experiments, this work employs five-fold cross-validation[47] for dataset partitioning. In this process, positive samples are composed of the observed edges, while negative samples are randomly selected from unconnected node pairs, ensuring a balanced sample set by matching the number of negative samples to the number of observed edges.

For link prediction, the fundamental task is to ensure that among all possible node pairs, the similarity scores assigned to existing edges are consistently higher than the scores assigned to non-existent edges.

## 3.2 Evaluation Metric

To evaluate the performance of the proposed model, we selects two key metrics: the Area Under the Curve (AUC)[48] and Average Precision (AP)[49][50]. Recognized as the most widely used evaluation standards in the fields of statistical physics and computer science, AUC and AP comprehensively reflect the model's predictive ability across different thresholds. Their specific calculations are implemented using functions provided by the standard Python library, scikit-learn [51]. Specifically, AUC focuses on the global ranking quality, assessing the overall capability of the model to distinguish positive samples from negative ones. AUC is a highly intuitive and stable metric when the ratio of positive and negative samples is relatively balanced. In contrast, AP emphasizes the prediction quality among top-ranked samples. Consequently, AP is considered more sensitive and practical than AUC in scenarios involving sample imbalance. The combination of these two metrics allows for a more comprehensive evaluation of the model's overall performance.

1、AUC: AUC measures the model's overall discriminative ability between positive and negative samples across all classification thresholds, and its value is equivalent to the area under the Receiver Operating Characteristic (ROC) curve:

$$AUC = \int_0^1 \text{TPR}(x)\,d(\text{FPR}(x)) \tag{1}$$

The ROC curve plots the relationship between the True Positive Rate (TPR) and the False Positive Rate (FPR) at various classification thresholds:

$$TPR = \frac{TP}{TP + FN} \tag{2}$$

$$FPR = \frac{FP}{FP + TN} \tag{3}$$

Here, TP represents the number of positive samples correctly predicted as positive (True Positives), FP is the number of negative samples incorrectly predicted as positive (False Positives), FN is the number of positive samples incorrectly predicted as negative (False Negatives), and TN represents the number of negative samples correctly predicted as negative (True Negatives).

Specifically, AUC can be interpreted as the probability that, when randomly selecting one existing link and one non-existent link from the test set, the model assigns a higher score to the former than to the latter. An AUC value closer to 1 signifies a stronger capability of the model to distinguish between positive and negative samples.

2、AP: AP is utilized to characterize the balance between Precision and Recall across different thresholds. It is equivalent to the area under the Precision-Recall (PR) curve. AP can be formally defined as:

$$AP = \sum_{n=1}^{N}(R_n - R_{n-1}) \cdot P_n \tag{4}$$

Here, $P_n$ and $R_n$ denote the model's Precision and Recall at the *n* threshold, respectively. Their calculation formulas are defined as:

$$P = \frac{TP}{TP + FP} \tag{5}$$

$$R = \frac{TP}{TP + FN} \tag{6}$$

Precision measures the proportion of true positive samples among all samples predicted as positive by the model. It can be defined as the ratio of edges that belong to the positive test set within the top L edges having the highest prediction scores. Average Precision (AP) provides a more comprehensive reflection of the model's performance in identifying positive samples by calculating the weighted average of Precision across different Recall levels. A higher AP value indicates that the model is better capable of ranking its correctly predicted edges at higher score positions. This suggests that the model maintains a better balance between Precision and Recall across different thresholds, thereby exhibiting more stable predictive performance.

### 3.3 Model overview

TELP integrates topological structure features and node embeddings, aiming to accurately predict missing or potential connections in the network. This model adheres to the supervised learning paradigm, reformulating the link prediction task as a binary classification problem.

Firstly, as illustrated in Figure 2, information is extracted from the real-world network to construct the sample set, followed by label assignment, where '1' denotes a positive sample and '0' denotes a negative sample. Next, the feature engineering module is completed in two steps: the screening and selection of structural topological features, and node feature embedding.

On the one hand, as shown in Figure 2(c), appropriate structural topological features are screened and selected based on the network type. Specifically, homogeneous networks utilize Common Neighbors (CN), Jaccard coefficient, Adamic–Adar index, and Resource Allocation (RA) index, while heterogeneous networks employ the Degree Heterogeneity Index (HI) to capture the local connection patterns of node pairs. On the other hand, as shown in Figure 2(d), low-dimensional embedding vectors are generated for each node using Node2Vec, which further encodes the global structural information of the network. Subsequently, the topological features are combined with the node embedding vectors to form the multi-dimensional feature vector used for model training.

Finally, as depicted in Figure 2(e), the link probability prediction module employs Logistic Regression, Random Forest, and XGBoost as base learners. TELP then utilizes soft voting to integrate the prediction results from these different models, thereby achieving the probability estimation of potential links between nodes. The performance of this model is evaluated under five-fold cross-validation to ensure the robustness and reliability of the prediction results.

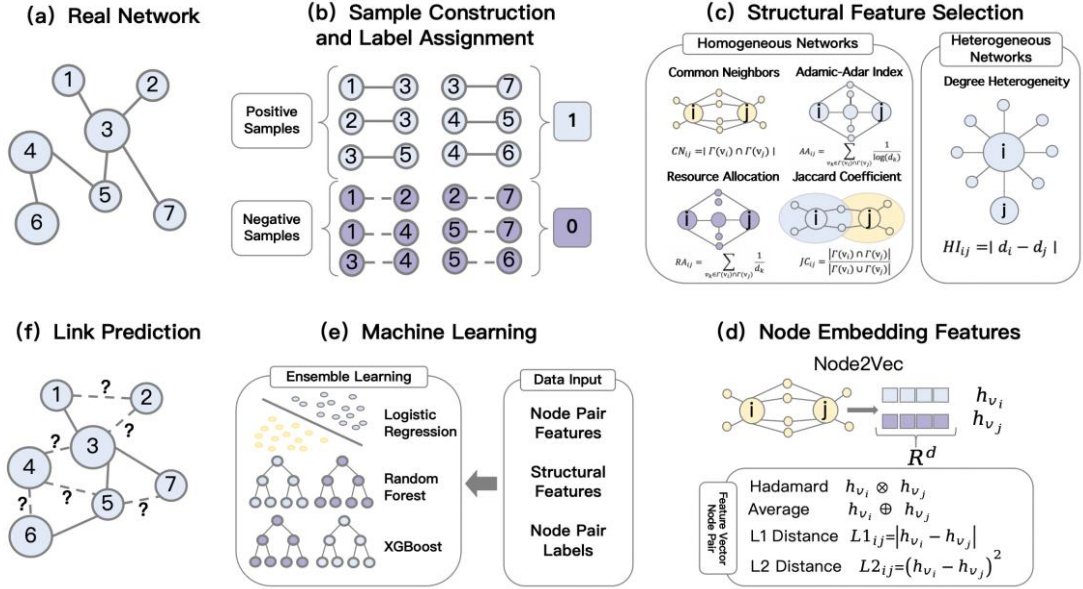

Figure 2. Feature construction and prediction workflow of the TELP model. *(a) Real network structure: The observed network serves as the basis, where nodes and existing links form the original topology. (b) Sample construction and label assignment: Existing links in the real network are extracted as positive samples, while negative samples are obtained by sampling node pairs from the set of non-links. Labels are then assigned to candidate node pairs, with existing links labeled as 1 and non-links labeled as 0. (c) Structural feature selection: For homogeneous networks, common local structural similarity indices—such as Common Neighbors, Adamic–Adar index, Resource Allocation index, and Jaccard coefficient—are extracted. For heterogeneous networks, degree heterogeneity is computed to capture structural differences between nodes. (d) Node embedding feature construction: Node2Vec is applied to the training subgraph through biased random walks and Skip-gram training to obtain d-dimensional embedding vectors. These embeddings are then combined into node-pair features using Hadamard product, averaging, L1 distance, and L2 distance. (e) Machine learning model training: Structural features and node embedding features are integrated into node-pair feature vectors and, together with corresponding labels, are fed into the ensemble learning model, which includes logistic regression, random forest, and XGBoost classifiers. (f) Link prediction: The trained model assigns scores to unobserved node pairs, producing link formation probabilities to predict missing or potential links within the network.*

## 3.4 Feature Engineering

### 3.4.1 Structural Topological Features

Structural topological features originate directly from the local connection patterns of the graph, enabling the efficient capture of connection structures between node pairs. Within this model, we classify networks into two distinct types—homogeneous networks and heterogeneous networks—and select the corresponding structural topological features to better adapt to their intrinsic connection patterns.

Previous relevant studies have demonstrated that the intrinsic structural properties of a network have a critical impact on the effectiveness of link prediction algorithms[14]. In homogeneous networks, nodes tend to connect with those exhibiting similar characteristics or residing within the same community; these networks typically have a uniform degree distribution with no distinct hub nodes. Consequently, the link prediction task for this type of network is usually based on the theory of triadic closure. Conversely, heterogeneous networks are characterized by significant differences in node degrees and the presence of a few high-degree hub nodes, resulting in a non-uniform overall degree distribution. In such cases, simple triadic closure indices are ineffective at capturing their unique connection patterns.

Therefore, we introduce four classic homogeneity indices to capture the topological structural features of homogeneous networks: the Common Neighbors index (CN)[1], the Jaccard coefficient [24], the Adamic–Adar index[23], and the Resource Allocation index (RA)[26]. For heterogeneous networks, we select the Degree Heterogeneity Index (HI)[8] as the core structural feature.

1. **Common Neighbors Index (CN):** The Common Neighbors Index (CN)[1] is based on the principle of triadic closure, positing that a higher number of shared neighbors implies a greater likelihood of connection. For any two nodes $v_i$ and $v_j$, the number of common neighbors is calculated by the following formula:

$$CN_{ij} = |\Gamma(v_i) \cap \Gamma(v_j)| \tag{7}$$

   Where $\Gamma(v_i)$ and $\Gamma(v_j)$ denote the sets of neighbors of node $v_i$ and node $v_j$, respectively.

2. **Jaccard Coefficient (JC):** The Jaccard Coefficient (JC) [24] serves as a normalized version of the CN index. It measures the proportion of common neighbors relative to the total number of neighbors. The calculation formula is:

$$JC_{ij} = \frac{|\Gamma(v_i) \cap \Gamma(v_j)|}{|\Gamma(v_i) \cup \Gamma(v_j)|} \tag{8}$$

3. **Adamic–Adar Index (AA):** The Adamic–Adar Index (AA) [23] extends the CN algorithm by applying a penalty based on the degrees of the common neighbors. This means that common neighbors with lower degrees contribute more significantly to the prediction score. Its

calculation formula is:

$$AA_{ij} = \sum_{v_k \in \Gamma(v_i) \cap \Gamma(v_j)} \frac{1}{\log(d_k)} \tag{9}$$

where $d_k$ is the degree of the common neighbor $v_k$.

4. **Resource Allocation Index (RA):** The Resource Allocation (RA) Index[26] simulates a resource allocation process across nodes in the network. When two nodes share a common neighbor, the common neighbor is viewed as allocating resources to that node pair. The lower the degree of the common neighbor, the higher the proportion of resources it can allocate to the corresponding node pair, and thus the greater its predictive contribution to the formation of a link between the pair. Its calculation formula is:

$$RA_{ij} = \sum_{v_k \in \Gamma(v_i) \cap \Gamma(v_j)} \frac{1}{d_k} \tag{10}$$

5. **Degree Heterogeneity Index (HI):** The Degree Heterogeneity Index (HI)[8] quantifies the connection heterogeneity between node pairs, reflecting the non-uniformity of the connection structure in the network. This index measures this metric using the absolute difference in the node degrees, and its definition is as follows:

$$HI_{ij} = | d_i - d_j | \tag{11}$$

**3.4.2 Node Embedding Features**

To capture the global topological information and deeper structural characteristics of the graph, our model employs the Node2Vec algorithm to generate low-dimensional, dense embedding vectors for every node in the graph. Node2Vec generates node sequences through biased random walk sampling and then utilizes the Skip-gram model to learn the vector representation of the nodes.The core strength of this method lies in its capability to encode the complex topological structure of the graph into a continuous vector space, ensuring that nodes structurally similar or belonging to the same community in the original network possess proximate representations in the vector space. The random walk process in Node2Vec is governed by the return parameter *p* and the in-out parameter *q*, which respectively regulate the probability of returning to the previously visited node and the tendency to explore new neighbors. This mechanism facilitates a balance between modeling local and global structural information.In our experiments, we systematically set and tuned the embedding dimension, walk length, number of walks, and the parameter combinations of *(p, q)* to thoroughly investigate their impact on link prediction performance. Crucially, in each fold of the cross-validation, our model runs the Node2Vec algorithm exclusively on the training subgraph, ensuring that the feature calculation process does not introduce any information from the test set.

After obtaining the node embedding vectors, the model proceeds to construct node-pair feature vectors to serve as the input for the machine learning classifier. For any candidate node pair $(v_i, v_j)$, twe retrieve their embeddings $h_{v_i}$ and node $h_{v_j}$ from the embedding lookup table produced by Node2Vec. These vectors are then merged into a single resultant vector based on a combination function. The proposed model incorporates four distinct combination methods to capture different types of interaction information between the node vectors. Let $h_{v_i}$ and $h_{v_j}$ be the $d$-dimensional embedding vectors for nodes $v_i$ and $v_j$, $h_{v_i}, h_{v_j} \in R^d$.

The four combination methods are as follows:

1. **Hadamard Product:** The Hadamard Product is an element-wise multiplication operation that captures the interplay between different dimensions of the node vectors. This method assumes that the relationship between nodes is a product of the independent interactions of individual dimensions, effectively preserving information from the original vectors. Its calculation formula is:

$$h_{v_i} \otimes h_{v_j} = \left[ h_{v_i}^{(1)} \cdot h_{v_j}^{(1)}, h_{v_i}^{(2)} \cdot h_{v_j}^{(2)}, \ldots \ldots, h_{v_i}^{(d)} \cdot h_{v_j}^{(d)} \right] \qquad (12)$$

2. **Average:** The Average method takes the mean of the two node embedding vectors, yielding a representative vector positioned between the two. Its calculation formula is:

$$h_{v_i} \oplus h_{v_j} = \frac{h_{v_i} + h_{v_j}}{2} \qquad (13)$$

3. **L1 Distance:** The L1 Distance measures the similarity between the two vectors by calculating the absolute difference of their corresponding elements. This method specifically focuses on quantifying the degree of dissimilarity between the two node embedding vectors. Its calculation formula is:

$$L1_{ij} = \left| h_{v_i} - h_{v_j} \right| = \left[ \left| h_{v_i}^{(1)} - h_{v_j}^{(1)} \right|, \ldots \ldots, \left| h_{v_i}^{(d)} - h_{v_j}^{(d)} \right| \right] \qquad (14)$$

4. **L2 Distance:** The L2 Distance measures the similarity by calculating the squared difference of their corresponding elements. This approach assigns greater weight to larger differences between the vectors. Its calculation formula is:

$$L1_{ij} = \left( h_{v_i} - h_{v_j} \right)^2 = \left[ \left( h_{v_i}^{(1)} - h_{v_j}^{(1)} \right)^2, \ldots \ldots, \left( h_{v_i}^{(d)} - h_{v_j}^{(d)} \right)^2 \right] \qquad (15)$$

Finally, we concatenate the network structural topological features with the combined node embedding vector to form a complete, multi-dimensional feature vector, which serves as the input data for the machine learning model. This modular design enables our model to flexibly select the optimal feature combination method tailored to the specific characteristics of different datasets.

### 3.5 Machine Learning

After obtaining the node-pair feature vectors, we treats the link prediction task as a binary classification problem and models it using supervised learning. To fully leverage the advantages of different models and enhance prediction stability, we construct an ensemble learning model. This model employs Logistic Regression[26], Random Forest[27], and XGBoost[28] as base learners, with each model independently fitted on the training set to output the predicted probability of the sample belonging to the positive class.

The base learners are chosen for their complementary predictive capabilities: Logistic Regression effectively captures the linear relationship between features and classes, offering inherent interpretability; Random Forest, conversely, utilizes voting across multiple decision trees to handle non-linear features and enhance generalization ability; while XGBoost iteratively optimizes model performance through a gradient boosting strategy, making it highly suitable for high-dimensional and complex feature combinations.

The ensemble learning component adopts a soft voting[52] strategy, which averages the probabilities output by the base learners to yield the final prediction result. Compared to a single model, soft voting allows for a more nuanced fusion of different models' prediction patterns, leading to superior performance.

We utilize five-fold cross-validation in the model training to ensure the robustness of the performance evaluation. In each fold of the cross-validation, the positive samples for the training set are derived from the training subgraph of the original network. Negative samples are generated via random sampling, maintaining an equal number of positive and negative samples, and ensuring no overlap between the training and test negative samples. Finally, the results from the five-fold cross-validation are averaged to obtain the overall performance of each single model and the ensemble model. This design not only guarantees fairness in training and evaluation but also facilitates the comparison of predictive effectiveness across different models and feature combinations, providing reliable quantitative support for link prediction.

### 4 Experiment

#### 4.1 Datasets

#### 4.1.1 Experimental Datasets

To comprehensively validate the effectiveness and generalization capability of our proposed link prediction model, we conduct experiments on nine large-scale public graph datasets spanning across nine different domains.

The Collaboration and Citation Networks category includes five networks: Netscience[53], CS[54], Cora[55], Citeseer[56], and DBLP[56]. The Netscience dataset describes a collaboration network among scholars engaged in network science research; nodes represent scholars, and an edge denotes collaboration if two scholars co-authored a paper. This network was compiled from the references of two review articles (M. Newman (2003) and S. Boccaletti et al. (2006)). Isolated nodes were removed prior to its use in this work. The CS dataset is a co-authorship network where each node represents an author, and an edge exists between two authors if they have jointly written at least one paper. Cora, Citeseer, and DBLP are classic citation networks, where nodes represent papers and edges indicate citation relationships.

The E-commerce Co-purchase Networks category contains two datasets: Photo[57] and Computers [58]. These datasets originate from product co-purchase behaviors on the Amazon e-commerce platform, reflecting association patterns between products. When users simultaneously purchase multiple products on the platform, potential association relationships are formed between these items. The graph structure—where nodes are products and edges represent co-purchase relationships—is constructed by mining and organizing large volumes of such user behavior data, thereby characterizing product association patterns.

The Social Networks category includes the Facebook[58] network. This network is a social relationship graph built upon the Facebook social platform, where nodes represent users and edges signify mutual social connection relationships.

Finally, ChChmn[59] is a Drug Interaction Network, where nodes represent drugs approved by the U.S. Food and Drug Administration (U.S. FDA), and edges indicate interactions between the drugs.

The statistical information for all datasets used in this work is presented in Table 1.

Table 1: Basic statistic of our dateset

| Dateset | Number of Nodes | Number of Edges | Maximum Degree | Minimum Degree | Average Degree | Global Clustering Coefficient |
|---|---|---|---|---|---|---|
| Netscience | 1461 | 2742 | 34 | 1 | 3.7536 | 0.693 |
| Facebook | 4039 | 88234 | 1045 | 1 | 43.6910 | 0.519 |
| ChChmn | 1514 | 48514 | 443 | 1 | 64.0872 | 0.259 |
| Photo | 7535 | 119081 | 1434 | 1 | 31.6074 | 0.177 |
| CS | 18333 | 81894 | 136 | 1 | 8.9341 | 0.183 |
| Computer | 13471 | 245861 | 2992 | 1 | 36.5023 | 0.108 |
| DBLP | 17716 | 52867 | 339 | 1 | 5.9683 | 0.134 |
| Cora | 2708 | 5278 | 168 | 1 | 3.8981 | 0.093 |
| Citeseer | 3312 | 4660 | 99 | 1 | 2.8140 | 0.130 |

### 4.1.2 Dataset Partitioning and Positive/Negative Sample Construction

To ensure the robustness and reliability of the model evaluation results, we adopts the five-fold cross-validation strategy. This method effectively reduces accidental errors caused by a single data partition by dividing the dataset into five equal-sized subsets, allowing each subset to serve as part of both the training and testing sets across different experimental folds, thereby guaranteeing the comprehensiveness and accuracy of the evaluation process.

In terms of dataset partitioning, we first randomly and non-overlappingly divide all genuinely existing edges in the dataset into five subsets of consistent size. In each fold of the experiment, one of these subsets is selected to serve as the positive samples for the test set, while the remaining four subsets are used to construct the training set. This partitioning approach guarantees that the positive samples in the test set have been completely unlearned by the model during the training phase, thus preventing cross-contamination between the training and testing sets and ensuring the fairness and rigor of the model evaluation.

For the construction of negative samples, we employ a negative sampling strategy to create a balanced binary classification task. During the training and testing phases of each fold, negative samples are generated by randomly selecting unconnected node pairs within the graph. These unconnected node pairs, along with their corresponding positive samples in the training set and test set, collectively form the balanced datasets used for model training and evaluation. To ensure the quantity of positive and negative samples is balanced in both the training and testing sets, we strictly control the number of negative samples in each fold, ensuring it equals the number of positive samples, thereby mitigating the negative impact that class imbalance could impose on model training and evaluation.

### 4.1.3  Network Type Classification

The Global Clustering Coefficient is a crucial topological metric for quantifying the degree of node aggregation in a network. It is defined as the ratio of the number of existing triadic closure structures to the number of all possible structures that could form a triad in the network. Essentially, it measures the actual proportion of triadic closure structures within the network, and its calculation formula is:

$$G_C = \frac{3 \cdot T}{N_\Delta} \tag{16}$$

where $G_C \in [0,1]$ is the Global Clustering Coefficient, $T$ is the number of actual triangles present in the network, and $N_\Delta$ is the total number of triplets in the network. Based on this metric, we classify the datasets into two major categories: homogeneous networks and heterogeneous networks. Existing research[14] suggests that $G_C = 0.2$ serves as a suitable demarcation threshold[9]. Accordingly, if the network's Global Clustering Coefficient is greater than or equal to 0.2, we treat it as a homogeneous network exhibiting significant triadic closure characteristics. Conversely, if the Global Clustering Coefficient is less than 0.2, we consider the network's connection pattern to display a high degree of non-uniformity and classify it as a heterogeneous network.

### 4.2  Experimental settings

In implementing the various baseline models, we adopt the parameter configurations recommended in their original implementations. To ensure both the optimal performance of our model and the reproducibility of the experimental results, we systematically configured the key hyperparameters for the Node2Vec algorithm and the machine learning models. The final experimental results presented are the averaged means from five-fold cross-validation, repeated over 10 independent runs.

**4.2.1 Node2Vec Parameters**

The parameter settings for the Node2Vec algorithm dictate how the random walk process explores the network structure, thereby influencing the learned embedding vectors. Table2 presents the specific parameter configuration used for Node2Vec.

1. **Embedding Dimension ($d$):** We select $d=64$ as the node embedding dimension. This dimension is chosen because it is sufficient to capture rich network information while avoiding the computational complexity associated with excessively high dimensionality.

2. **Walk Length ($l$):** This parameter defines the number of steps in each random walk. A longer walk length helps in capturing relationships between more distant nodes.

3. **Number of Walks ($n$):** Multiple random walks starting from each node generate a sufficiently diverse set of node sequences, providing rich contextual information for the subsequent Skip-gram model training and ensuring the learning of robust node representations.

4. **Biased Random Walk ($p, q$):** The core of Node2Vec lies in its biased random walk, controlled by the return parameter ($p$) and the in-out parameter ($q$). A setting of $p > 1$ biases the walk to stay away from the most recently visited node, encouraging broader exploration, while $q < 1$ biases the walk to explore distant nodes, promoting exploration beyond the immediate neighborhood.

**4.2.2 Embedding Combination Strategy**

When combining the embedding vectors of two nodes to form the feature vector for an edge, we tested multiple combination methods, including the Hadamard Product, Average, L1 Distance, and L2 Distance. Through extensive preliminary experiments conducted across various datasets, we determined that the L1 Distance combination strategy achieved superior link prediction performance in the majority of networks. Consequently, the L1 Distance is uniformly adopted as the method for combining embedding vectors throughout all experiments.

Furthermore, we unified the basic parameters such as embedding dimension ($d$), walk length ($l$), and number of walks ($n$), and specifically adjusted the Node2Vec parameters $p$ and $q$ to characterize the structural differences between homogeneous and heterogeneous networks. This tailored approach allows for a better reflection of the differentiated features relevant to link prediction in various network types.

We conducted systematic preliminary experiments on representative datasets such as Netscience, Cora, and Photo, setting the following ranges for key parameters: $d \in \{32, 64, 128\}$、

$l\in\{20,30,40,50,60,70\}$、$n\in\{200,300,400,500,800\}$. The results indicated that: when $d$ was too small, the embedding's representational capacity was insufficient, leading to a noticeable decrease in the AUC metric on complex networks; conversely, when $d$ was too large, the training time and computational complexity increased significantly. A short $l$ restricted the local structure covered by the random walk, preventing the embeddings from fully capturing structural information; while a long $l$ introduced excessive noise edges in certain networks, negatively impacting embedding quality. When the number of random walks ($n$) was too small, the embeddings were unstable; when too large, the computational overhead was prohibitive.

Based on a comprehensive consideration of embedding quality, algorithm generalization ability, and stability across different datasets, we ultimately set $d$=64, $l$=40, and $n$=400 uniformly for all experiments. This standardized setting guarantees the expressive power of the embeddings while controlling training costs and preventing performance bias resulting from parameter scale differences, thereby ensuring fair and stable experimental comparisons between different networks.

Regarding the control parameters for Node2Vec's random walk bias, $p$ and $q$, we set them based on their influence principles on the walk strategy: Parameter $p$ controls the probability of returning to the previous node, with smaller values favoring Depth-First Search (DFS); Parameter $q$ controls the tendency to explore new nodes, with smaller values favoring Breadth-First Search (BFS), and larger values favoring DFS. In preliminary experiments, we systematically tested the ranges $p\in[0.1,3]$ and $q\in[0.1,3]$. The results showed that for highly homogeneous networks, smaller $p$ and larger $q$ effectively reinforce local neighborhood sampling, and the majority of datasets achieved optimal link prediction performance under the L1 combination. For highly heterogeneous networks, the complementary setting helps capture cross-community and long-range dependency structures. Based on these findings, we ultimately selected two stable and well-performing parameter configurations: $p$=1, $q$=1.8 for homogeneous networks and $p$=3, $q$=0.3 for heterogeneous networks.

Table 2  Experimental Parameters

| Dataset | $d$ | $l$ | $n$ | $p$ | $q$ | Combination Method |
|---|---|---|---|---|---|---|
| Netscience | 64 | 40 | 400 | 1 | 1.8 | L1 |
| Facebook | 64 | 40 | 400 | 1 | 1.8 | L1 |
| ChChmn | 64 | 40 | 400 | 1 | 1.8 | L1 |
| Photo | 64 | 40 | 400 | 3 | 0.3 | L1 |
| CS | 64 | 40 | 400 | 3 | 0.3 | L1 |
| Computer | 64 | 40 | 400 | 3 | 0.3 | L1 |
| DBLP | 64 | 40 | 400 | 3 | 0.3 | L1 |
| Cora | 64 | 40 | 400 | 3 | 0.3 | L1 |
| Citeseer | 64 | 40 | 400 | 3 | 0.3 | L1 |

**4.3 Performance Comparison**

We contrast TELP with two major categories of widely used baseline models: (1) Traditional link prediction methods, including the Common Neighbors (CN) index, Jaccard Coefficient, Adamic–

Adar index, Resource Allocation (RA) index, and the Degree Heterogeneity Index (HI). (2) Current mainstream learning-based Graph Neural Network (GNN) models, specifically the Graph Convolutional Network (GCN), Graph Sample and Aggregate Network (GraphSAGE)[60], and Graph Attention Network (GAT)[61].As demonstrated in Table 3 (AUC) and Table 4 (AP), TELP consistently outperforms all baseline methods across all nine datasets, exhibiting stable and superior overall performance.

Compared to traditional link prediction methods, TELP achieved a significant performance improvement across all datasets. In the three homogeneous networks Netscience, Facebook and ChChmn, the AUC and AP scores of the CN, JC, AA and RA algorithms remained relatively stable and generally exceeded 0.83. Specifically, CN and AA consistently achieved AUC scores above 0.9. However, the AUC and AP scores obtained by TELP consistently exceeded 0.93. Furthermore, even for the Facebook network, where the CN, JC, AA, and RA methods scored highly in the 0.987–0.993 range for both AUC and AP, the proposed model's scores remained stably higher than all four. Conversely, the HI algorithm performed poorly in homogeneous networks, with its AUC score in the Netscience network even dropping below 0.5. This result further validates the necessity of differentiating feature selection based on network structure.

In heterogeneous networks, the performance of traditional link prediction algorithms varied significantly. CN, JC, AA, and RA performed relatively stable in the Photo, CS, and Computers networks, with AUC and AP scores ranging between 0.882 and 0.968. However, their performance was notably weaker in the DBLP, Cora, and Citeseer networks, where both AUC and AP scores fell within the 0.65–0.765 range. Nevertheless, the AUC and AP scores achieved by TELP remained stably superior to all traditional algorithms across all six heterogeneous networks, reaching 0.984 for AUC and 0.982 for AP on the Photo network. The HI algorithm showed improved performance in heterogeneous networks compared to its results on homogeneous ones, but its AUC and AP scores remained stably below 0.85.These results collectively demonstrate that incorporating node embedding features and classifying network types for targeted feature capture effectively enhances predictive accuracy.

In comparison with the learning-based Graph Neural Network (GNN) models, our model maintains a stable advantage across all datasets. In homogeneous networks, with the exception of the Facebook network, the AUC and AP scores achieved by GCN, GAT, and GraphSAGE were all below 0.9. In contrast, the scores obtained by our model demonstrated a distinct advantage.In heterogeneous networks, the fundamental Graph Neural Networks still exhibited moderate performance. Among these, GraphSAGE achieved the highest AUC and AP scores, consistently exceeding 0.8 across the six networks, yet these scores were still stably lower than the results produced by TELP.

Table 3: AUC Results of Algorithms on Datasets

| | **AUC** | Netscience | Facebook | ChChmn | Photo | CS | Computers | DBLP | Cora | Citeseer |
|---|---|---|---|---|---|---|---|---|---|---|
| **Heuristic Methods** | CN | 0.908 | 0.990 | 0.904 | 0.963 | 0.882 | 0.959 | 0.762 | 0.700 | 0.656 |
| | JC | 0.908 | 0.988 | 0.881 | 0.956 | 0.882 | 0.948 | 0.762 | 0.700 | 0.657 |
| | AA | 0.908 | 0.992 | 0.906 | 0.966 | 0.882 | 0.963 | 0.762 | 0.700 | 0.656 |
| | RA | 0.909 | 0.993 | 0.907 | 0.968 | 0.883 | 0.966 | 0.762 | 0.701 | 0.657 |
| | HI | 0.464 | 0.497 | 0.627 | 0.671 | 0.627 | 0.709 | 0.690 | 0.662 | 0.605 |
| **Neural Network Methods** | GCN | 0.789 | 0.848 | 0.843 | 0.795 | 0.695 | 0.812 | 0.730 | 0.679 | 0.626 |
| | GAT | 0.842 | 0.900 | 0.777 | 0.660 | 0.731 | 0.794 | 0.786 | 0.663 | 0.703 |
| | GraphSAGE | 0.865 | 0.903 | 0.895 | 0.859 | 0.809 | 0.880 | 0.866 | 0.803 | 0.806 |
| **TELP** | LR | **0.941** | **0.995** | **0.936** | **0.983** | **0.957** | **0.980** | **0.909** | **0.879** | **0.818** |
| | RF | **0.937** | **0.994** | **0.951** | **0.982** | **0.951** | **0.978** | **0.900** | **0.867** | **0.800** |
| | XGB | **0.938** | **0.995** | **0.958** | **0.983** | **0.954** | **0.980** | **0.908** | **0.875** | **0.816** |
| | SV | **0.938** | **0.995** | **0.956** | **0.984** | **0.954** | **0.980** | **0.902** | **0.873** | **0.807** |

Table 4: AP Results of Algorithms on Datasets

| | **AP** | Netscience | Facebook | ChChmn | Photo | CS | Computers | DBLP | Cora | Citeseer |
|---|---|---|---|---|---|---|---|---|---|---|
| **Heuristic Methods** | CN | 0.908 | 0.988 | 0.891 | 0.960 | 0.882 | 0.954 | 0.762 | 0.700 | 0.657 |
| | JC | 0.908 | 0.987 | 0.837 | 0.951 | 0.882 | 0.942 | 0.761 | 0.697 | 0.656 |
| | AA | 0.909 | 0.991 | 0.895 | 0.966 | 0.883 | 0.964 | 0.763 | 0.702 | 0.656 |
| | RA | 0.909 | 0.993 | 0.894 | 0.968 | 0.883 | 0.967 | 0.762 | 0.702 | 0.657 |
| | HI | 0.512 | 0.513 | 0.610 | 0.716 | 0.635 | 0.754 | 0.702 | 0.692 | 0.625 |
| **Neural Network Methods** | GCN | 0.786 | 0.848 | 0.809 | 0.792 | 0.712 | 0.818 | 0.750 | 0.708 | 0.653 |
| | GAT | 0.846 | 0.896 | 0.715 | 0.633 | 0.722 | 0.776 | 0.793 | 0.673 | 0.729 |
| | GraphSAGE | 0.865 | 0.900 | 0.879 | 0.861 | 0.810 | 0.884 | 0.872 | 0.804 | 0.814 |
| **TELP** | LR | **0.962** | **0.994** | **0.930** | **0.981** | **0.966** | **0.978** | **0.916** | **0.899** | **0.854** |
| | RF | **0.955** | **0.992** | **0.949** | **0.979** | **0.962** | **0.976** | **0.922** | **0.892** | **0.844** |
| | XGB | **0.959** | **0.994** | **0.957** | **0.982** | **0.966** | **0.979** | **0.932** | **0.902** | **0.861** |
| | SV | **0.960** | **0.994** | **0.956** | **0.982** | **0.966** | **0.979** | **0.928** | **0.900** | **0.854** |

### 4.4 Ablation Study

TELP is composed of three components: feature screening and extraction, node embedding, and ensemble learning. To further validate the performance gains contributed by the individual modules and the necessity of their combination within this link prediction model, we designed and conducted two sets of ablation studies.

Ablation Study I: involves removing the node embedding module while retaining the structural topological feature screening and ensemble learning modules. This aims to investigate the independent supporting role of structural topological features on algorithm predictive performance and their importance within the overall model.

Ablation Study II: nvolves removing the structural topological feature screening and extraction module while retaining the node embedding and ensemble learning modules. This is designed to clarify the representational capability of node embeddings for high-order structural information and to verify its contribution to the overall model performance.

Both sets of ablation experiments utilized the same datasets and experimental settings as the full model to ensure the comparability of the results.

**4.4.1 Ablation Study I: Retaining Only Structural Feature Screening and Ensemble Learning**

As the experimental results in Table 5 (AUC), Table 6 (AP), and Figure 3 (AUC), Figure 4 (AP) indicate, the removal of the node embedding module led to a significant decline in model performance across all nine datasets. Furthermore, the loss of the node embedding feature resulted in a substantially larger drop in prediction effectiveness within the heterogeneous networks compared to the homogeneous ones. Taking the SoftVoting ensemble model as an example, the AUC values in the large-scale heterogeneous networks (Photo, CS, and Computers) dropped notably, from 0.984, 0.954, and 0.980 to 0.672, 0.627, and 0.709, respectively, representing an average decrease of over 0.25. AP values also experienced a clear reduction; for instance, the Photo dataset's AP dropped from 0.982 to 0.716.

This significant reduction demonstrates that relying solely on local topological features, such as CN, Jaccard Coefficient, AA, RA, or HI, is effective for characterizing the neighborhood similarity of node pairs but insufficient for capturing potential high-order structural relationships and the network's global topological patterns. In real-world complex networks, local features alone often fail to adequately distinguish between node pairs that share similar local structure but possess differing global properties, thereby leading to a significant reduction in predictive accuracy. This result validates that the node embedding module provides crucial complementary information about the global latent structure and is a core source for enhancing predictive performance.

Table 5: AUC Comparison of Ablation Study I and the Full Model on Various Datasets

|  | AUC | Netscience | Facebook | ChChmn | Photo | CS | Computers | DBLP | Cora | Citeseer |
|---|---|---|---|---|---|---|---|---|---|---|
| **Ablation Study I** | LR | 0.909 | 0.991 | 0.908 | 0.671 | 0.628 | 0.709 | 0.691 | 0.662 | 0.603 |
|  | RF | 0.908 | 0.989 | 0.915 | 0.673 | 0.627 | 0.709 | 0.690 | 0.660 | 0.610 |
|  | XGB | 0.909 | 0.993 | 0.919 | 0.673 | 0.627 | 0.709 | 0.690 | 0.661 | 0.611 |
|  | SV | 0.909 | 0.992 | 0.921 | 0.672 | 0.627 | 0.709 | 0.690 | 0.662 | 0.612 |
| **TELP** | LR | **0.941** | **0.995** | **0.936** | **0.983** | **0.957** | **0.980** | **0.909** | **0.879** | **0.818** |
|  | RF | **0.937** | **0.994** | **0.951** | **0.982** | **0.951** | **0.978** | **0.900** | **0.867** | **0.800** |
|  | XGB | **0.938** | **0.995** | **0.958** | **0.983** | **0.954** | **0.980** | **0.908** | **0.875** | **0.816** |
|  | SV | **0.938** | **0.995** | **0.956** | **0.984** | **0.954** | **0.980** | **0.902** | **0.873** | **0.807** |

Table 6: AP Comparison of Ablation Study I and the Full Model on Various Datasets

|  | AP | Netscience | Facebook | ChChmn | Photo | CS | Computers | DBLP | Cora | Citeseer |
|---|---|---|---|---|---|---|---|---|---|---|
| **Ablation Study I** | LR | 0.909 | 0.991 | 0.899 | 0.715 | 0.636 | 0.753 | 0.703 | 0.690 | 0.624 |
|  | RF | 0.908 | 0.987 | 0.902 | 0.715 | 0.636 | 0.752 | 0.703 | 0.684 | 0.622 |
|  | XGB | 0.909 | 0.992 | 0.911 | 0.716 | 0.636 | 0.753 | 0.703 | 0.685 | 0.623 |
|  | SV | 0.909 | 0.992 | 0.923 | 0.716 | 0.636 | 0.753 | 0.703 | 0.689 | 0.626 |
| **TELP** | LR | **0.962** | **0.994** | **0.930** | **0.981** | **0.966** | **0.978** | **0.916** | **0.899** | **0.854** |
|  | RF | **0.955** | **0.992** | **0.949** | **0.979** | **0.962** | **0.976** | **0.922** | **0.892** | **0.844** |
|  | XGB | **0.959** | **0.994** | **0.957** | **0.982** | **0.966** | **0.979** | **0.932** | **0.902** | **0.861** |
|  | SV | **0.960** | **0.994** | **0.956** | **0.982** | **0.966** | **0.979** | **0.928** | **0.900** | **0.854** |

**4.4.2 Ablation Study II: Retaining Only Node Embedding and Ensemble Learning**

The experimental results, as presented in Table 7 (AUC), Table 8 (AP), and Figure 3 (AUC), Figure 4 (AP), indicate that after removing the structural topological feature screening and extraction, the overall model performance still maintained a high level, stably outperforming traditional baseline methods, yet it showed a slight decline compared to the full model. Taking the SoftVoting ensemble model as an example, the average difference in AUC between using embedding features alone and the complete model was approximately 0.0062 across the nine datasets; the average difference in AP was approximately 0.0071.

In the majority of datasets, such as Netscience, Facebook, and Photo, the predictive effectiveness of the embeddings alone closely approached that of the full model. This demonstrates that Node2Vec is capable of effectively encoding a substantial amount of discriminative high-order and global structural information, thus serving as the main source of the model's performance support.

However, on the ChChmn, Cora, and Citeseer datasets, using the embedding alone resulted in a relatively noticeable performance degradation compared to the full model, with AUC differences of approximately 0.018, 0.015, and 0.009, respectively; the corresponding AP differences were approximately 0.030, 0.011, and 0.005. This suggests that in these specific networks, explicit topological features still provide complementary information regarding local structure or degree-related patterns, which yields additional predictive gains.

Therefore, although the Node2Vec embedding can highly generalize the high-order structural information of nodes in most scenarios, explicit topological indices maintain an irreplaceable complementary value in networks with specific topological characteristics. This further proves the necessity and complementarity of the proposed design combining topological features and embeddings.

Table 7: AUC Comparison of Ablation Study II and the Full Model on Various Datasets

|  | AUC | Netscience | Facebook | ChChmn | Photo | CS | Computers | DBLP | Cora | Citeseer |
|---|---|---|---|---|---|---|---|---|---|---|
| **Ablation Study II** | LR | 0.939 | 0.992 | 0.923 | 0.981 | 0.957 | 0.976 | 0.907 | 0.837 | 0.814 |
|  | RF | 0.933 | 0.992 | 0.938 | 0.979 | 0.951 | 0.973 | 0.896 | 0.852 | 0.793 |
|  | XGB | 0.937 | 0.992 | 0.937 | 0.981 | 0.954 | 0.976 | 0.907 | 0.856 | 0.806 |
|  | SV | 0.935 | 0.992 | 0.938 | 0.981 | 0.953 | 0.977 | 0.901 | 0.858 | 0.798 |
| **TELP** | LR | **0.941** | **0.995** | **0.936** | **0.983** | **0.957** | **0.980** | **0.909** | **0.879** | **0.818** |
|  | RF | **0.937** | **0.994** | **0.951** | **0.982** | **0.951** | **0.978** | **0.900** | **0.867** | **0.800** |
|  | XGB | **0.938** | **0.995** | **0.958** | **0.983** | **0.954** | **0.980** | **0.908** | **0.875** | **0.816** |
|  | SV | **0.938** | **0.995** | **0.956** | **0.984** | **0.954** | **0.980** | **0.902** | **0.873** | **0.807** |

Table 8: AP Comparison of Ablation Study II and the Full Model on Various Datasets

|  | AP | Netscience | Facebook | ChChmn | Photo | CS | Computers | DBLP | Cora | Citeseer |
|---|---|---|---|---|---|---|---|---|---|---|
| **Ablation Study II** | LR | 0.951 | 0.990 | 0.899 | 0.979 | 0.966 | 0.974 | 0.914 | 0.891 | 0.850 |
|  | RF | 0.952 | 0.989 | 0.927 | 0.976 | 0.961 | 0.971 | 0.919 | 0.878 | 0.838 |
|  | XGB | 0.956 | 0.990 | 0.924 | 0.978 | 0.966 | 0.974 | 0.929 | 0.887 | 0.853 |
|  | SV | 0.955 | 0.991 | 0.926 | 0.979 | 0.965 | 0.975 | 0.926 | 0.889 | 0.849 |
| **TELP** | LR | **0.962** | **0.994** | **0.930** | **0.981** | **0.966** | **0.978** | **0.916** | **0.899** | **0.854** |
|  | RF | **0.955** | **0.992** | **0.949** | **0.979** | **0.962** | **0.976** | **0.922** | **0.892** | **0.844** |
|  | XGB | **0.959** | **0.994** | **0.957** | **0.982** | **0.966** | **0.979** | **0.932** | **0.902** | **0.861** |
|  | SV | **0.960** | **0.994** | **0.956** | **0.982** | **0.966** | **0.979** | **0.928** | **0.900** | **0.854** |

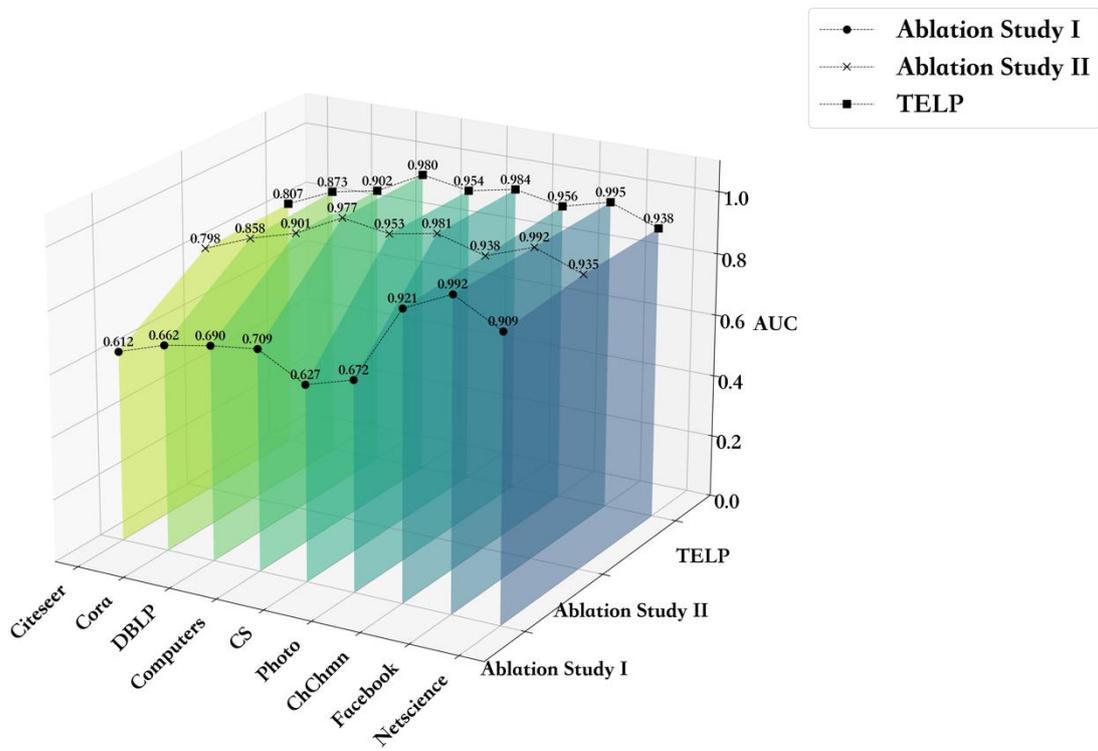

Figure 3: AUC Performance Comparison of Different Models Across Datasets

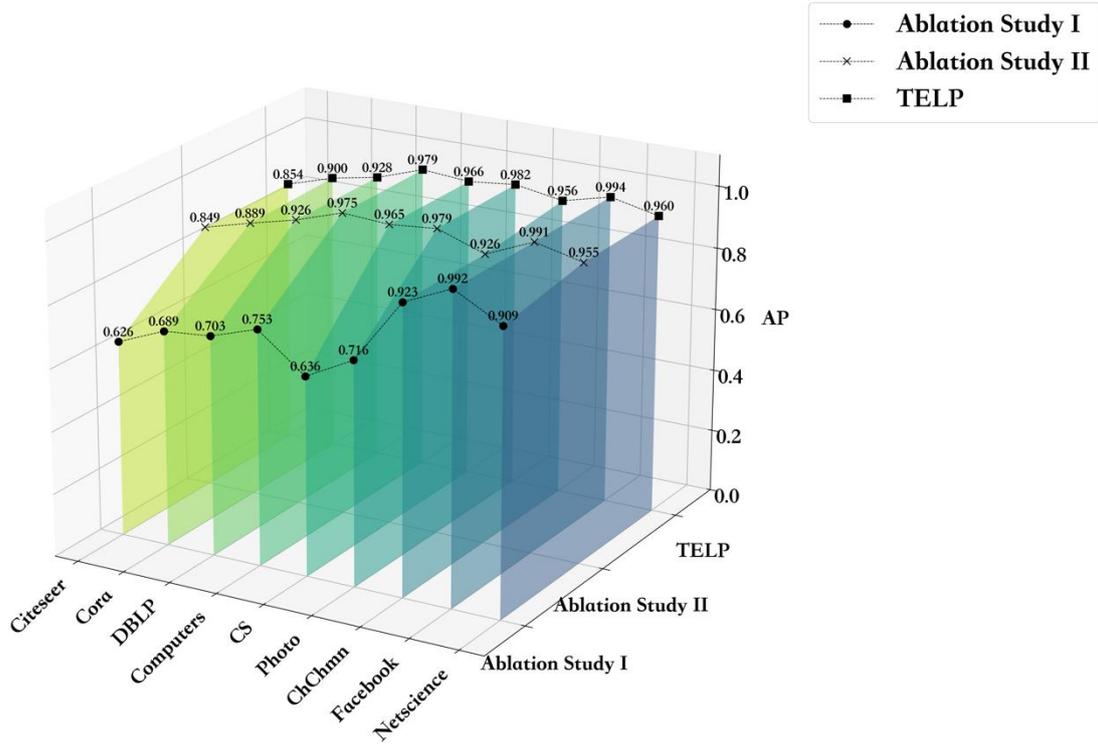

Figure 4: AP Performance Comparison of Different Models Across Datasets

By comparing the two ablation studies with the full model, we draw several key conclusions. First, using only the local structural topological features causes a significant decline in algorithm performance, which confirms the irreplaceable role of embedding features in representing the network's latent structural information. Second, although using only node embedding features yields a high overall performance, the comprehensive prediction results remain slightly inferior to the full model. This suggests that in certain networks, the structural topological features still provide crucial local supplementary information. Finally, when both types of features are combined, the model is able to simultaneously account for the interpretability of local structure and the sufficiency of global representation, thereby achieving the optimal link prediction effect across all nine datasets.

These results strongly support the underlying design rationale of TELP. The structural-topology feature extraction boosts local discrimination, the node embeddings encode higher-order relational structure, and the ensemble model fuses both representations to deliver superior predictive performance.

## 5 Conclusion

We propose an integrated link prediction model, TELP, which fuses topological features and embedding representations. This model aims to address the limitations of traditional heuristic and existing learning-based prediction algorithms, ensuring the algorithm maintains high efficiency and simplicity while balancing predictive accuracy and interpretability. The core idea of this model lies in differentially selecting topological indices based on network structural characteristics, fusing them with node embedding features, and ultimately utilizing an ensemble learning model for robust prediction. Experiments conducted on nine commonly used real-world datasets spanning citation, social, and e-commerce networks demonstrate that our model significantly outperforms both traditional heuristic algorithms and existing fundamental Graph Neural Network models in terms of AUC and AP. Furthermore, through ablation studies, we further confirmed that node embedding features provide global structural information that cannot be supplemented by local structural features in link prediction.

From a theoretical perspective, our research retains the core logic of network topological structure analysis rooted in statistical physics. By organically integrating traditional topological quantification with learnable global interaction models, the model effectively reveals the latent organizational mechanisms of complex networks. Unlike deep learning models, our approach utilizes graph embedding and classical machine learning methods to express and model topological features in a more transparent manner. This ensures that the algorithm's results correspond to the physical interpretation inherent in network science, thereby offering a novel computational perspective for the structural dynamics analysis of complex systems.

From an application perspective, the TELP model demonstrates excellent generality and scalability, capable of adapting to diverse network data structures across different fields. It provides a highly interpretable prediction tool with low computational cost for various multidisciplinary scenarios, including social science, bioinformatics, knowledge graph construction, and recommendation systems.

Despite the excellent and stable performance demonstrated by the proposed model, several avenues for further improvement still exist. Firstly, the model's feature engineering component relies on the Node2Vec algorithm, whose parameter settings exert a certain influence on the final performance; thus, future work can explore automated parameter tuning or the integration of novel graph embedding methods. Secondly, the current model is primarily designed for static networks. Thus, exploring how to extend it to dynamic networks to effectively cope with constantly evolving network structures represents a crucial direction for future research.

## 6 Data and codes

All cleaned data and the associated source code have been made publicly available in the GitHub repository at: https://github.com/Zi-XuanJin/TELP. This repository includes the complete implementation code, preprocessing scripts, and the cleaned dataset used in our experiments, facilitating reproducibility and further exploration. Researchers and developers are welcome to access, use, and contribute to the codebase.

**References**

[1] Liben-Nowell D, Kleinberg J. The link prediction problem for social networks//Proceedings of the 12th International Conference on Information and Knowledge Management (CIKM '03). 2003: 556-559.

[2] Ran Y J, Liu S Y, Yu X Y, Shang K K, Jia T. Predicting future links with new nodes in temporal academic networks. Journal of Physics: Complexity, 2022, 3(1): 015006.

[3] Yu H, et al. High-quality binary protein interaction map of the yeast interactome network. Science, 2008, 322: 104.

[4] Benchettara N, Kanawati R, Rouveirol C. A supervised machine learning link prediction approach for academic collaboration recommendation//Proceedings of the 4th ACM Conference on Recommender Systems. 2010: 253-256.

[5] Li X, Chen H. Recommendation as link prediction in bipartite graphs: a graph kernel-based machine learning approach. Decision Support Systems, 2013, 54(2): 880-890.

[6] Nickel M, et al. A review of relational machine learning for knowledge graphs. Proceedings of the IEEE, 2015, 104(1): 11-33.

[7] Lü L Y. Link prediction in complex networks. Journal of University of Electronic Science and Technology of China, 2010, 39(5): 651-661 (in Chinese)

[8] Shang K K, Li T C, Small M, Burton D, Wang Y. Link prediction for tree-like networks. Chaos: An Interdisciplinary Journal of Nonlinear Science, 2019, 29(6).

[9] Wang P, et al. Link prediction in social networks: the state-of-the-art. Science China Information Sciences, 2015, 58(1): 1-38.

[10] Al Hasan M, et al. Link prediction using supervised learning//SDM06 Workshop on Link Analysis, Counter-terrorism and Security. 2006: 30.

[11] Zhao Z L, et al. A novel link prediction algorithm based on inductive matrix completion. Expert Systems with Applications, 2022, 188: 116033.

[12] Lichtenwalter R N, Lussier J T, Chawla N V. New perspectives and methods in link prediction//Proceedings of the 16th ACM SIGKDD International Conference on Knowledge Discovery and Data Mining. 2010.

[13] Kapoor P, et al. A survey on feature extraction and learning techniques for link prediction in homogeneous and heterogeneous complex networks. Artificial Intelligence Review, 2024, 57(12): 348.

[14] Shang K K, et al. Triadic Closure-Heterogeneity-Harmony GCN for Link Prediction. arXiv preprint arXiv:2504.20492v1, 2025.

[15] Wu Z H, Pan S, Chen F W, Long G D, Zhang C Q, Yu P S. A comprehensive survey on graph neural networks. IEEE Transactions on Neural Networks and Learning Systems, 2020, 32(1): 4-24.

[16] Kipf T N, Welling M. Semi-supervised classification with graph convolutional networks//International Conference on Learning Representations (ICLR). 2017.

[17] Velickovic P, Cucurull G, Casanova A, Romero A, Lio P, Bengio Y. Graph attention networks//International Conference on Learning Representations (ICLR). 2018.

[18] Hamilton W, Ying Z, Leskovec J. Inductive representation learning on large graphs//Advances in Neural Information Processing Systems (NeurIPS 2017), Vol. 30. 2017.

[19] Zhang M H, Chen Y X. Link prediction based on graph neural networks. Advances in Neural Information Processing Systems, 2018, 31.

[20] Pósfai M, Barabási A L. Network Science. Vol. 3. Cambridge, UK: Cambridge University Press, 2016.

[21] Belkin M, Niyogi P. Laplacian eigenmaps and spectral techniques for embedding and clustering//Advances in Neural Information Processing Systems (NeurIPS 2001), Vol. 14. 2001.

[22] Jaccard P. Étude comparative de la distribution florale dans une portion des Alpes et des Jura. Bulletin de la Société Vaudoise des Sciences Naturelles, 1901, 37: 547-579.

[23] Adamic L A, Adar E. Friends and neighbors on the Web. Social Networks, 2003, 25(3): 211-230.

[24] Zhou T, Lü L, Zhang Y C. Predicting missing links via local information. Eur. Phys. J. B, 2009, 71: 623.

[25] Grover A, Leskovec J. node2vec: scalable feature learning for networks//Proceedings of the 22nd ACM SIGKDD International Conference on Knowledge Discovery and Data Mining. 2016: 855-864.

[26] Cox D R. The regression analysis of binary sequences. Journal of the Royal Statistical Society Series B: Statistical Methodology, 1958, 20(2): 215-232.

[27] Breiman L. Random forests. Machine Learning, 2001, 45(1): 5-32.

[28] Chen T, Guestrin C. XGBoost: a scalable tree boosting system//Proceedings of the 22nd ACM SIGKDD International Conference on Knowledge Discovery and Data Mining. 2016: 785-794.

[29] Salton G, McGill M J. Introduction to Modern Information Retrieval. Auckland: MuGraw-Hill, 1983.

[30] Sorensen T. A method of establishing groups of equal amplitude in plant sociology based on similarity of species content and its application to analyses of the vegetation on Danish commons. Biol. Skr., 1948, 5: 1.

[31] Lü L, Zhou T. Link prediction in weighted networks: the role of weak ties. EPL (Europhysics Letters), 2010, 89(1): 18001.

[32] Heckerman D, Geiger D, Chickering D. Learning bayesian networks: the combination of knowledge and statistical data. Machine Learning, 1995, 20: 197-243.

[33] Taskar B, Abbeel P, Koller D. Discriminative probabilistic models for relational data//Proc. UAI 2002. 2002: 485.

[34] Neville J, Jensen D. Relational dependency networks. Journal of Machine Learning Research, 2007, 8: 653-692.

[35] Clauset A, Moore C, Newman M E J. Hierarchical structure and the prediction of missing links in networks. Nature, 2008, 453: 98-101.

[36] Holland P W, Laskey K B, Leinhard S. Stochastic blockmodels: first steps. Social Networks, 1983, 5: 109-137.

[37] Lü L, Zhou T. Link prediction in complex networks: a survey. Physica A: Statistical Mechanics and its Applications, 2011, 390(6): 1150-1170.

[38] Fire M, et al. Link prediction in social networks using computationally efficient topological features//2011 IEEE Third International Conference on Privacy, Security, Risk and Trust & 2011 IEEE Third International Conference on Social Computing. IEEE, 2011.

[39] Ben-Hur A, Noble W S. Kernel methods for predicting protein–protein interactions. Bioinformatics, 2005, 21(suppl_1): i38-i46.

[40] Al Hasan M, Zaki M J. A survey of link prediction in social networks//Social Network Data Analytics. Springer US, 2011: 243-275.

[41] Perozzi B, Al-Rfou R, Skiena S. Deepwalk: online learning of social representations//Proceedings of the 20th ACM SIGKDD International Conference on Knowledge Discovery and Data Mining. 2014: 701-710.

[42] Grover A, Leskovec J. node2vec: scalable feature learning for networks//Proceedings of the 22nd ACM SIGKDD International Conference on Knowledge Discovery and Data Mining. 2016.

[43] Cao S, Lu W, Xu Q. GraRep: learning graph representations with global structural information//Proceedings of the 24th ACM International Conference on Information and Knowledge Management. 2015: 891-900.

[44] Ou M, et al. Asymmetric transitivity preserving graph embedding//Proceedings of the 22nd ACM SIGKDD International Conference on Knowledge Discovery and Data Mining. 2016: 1105-1170.

[45] Trouillon T, et al. Complex embeddings for simple link prediction//International Conference on Machine Learning (ICML). PMLR, 2016.

[46] Friedman J. The Elements of Statistical Learning: Data Mining, Inference, and Prediction. 2009.

[47] Kohavi R. A study of cross-validation and bootstrap for accuracy estimation and model selection//Proceedings of the 14th International Joint Conference on Artificial Intelligence - Volume 2 (IJCAI '95). Morgan Kaufmann Publishers Inc., San Francisco, CA, USA, 1995: 1137-1143.

[48] Hanley J A, McNeil B J. The meaning and use of the area under a receiver operating characteristic (ROC) curve. Radiology, 1982, 143(1).

[49] Aslam J A, Pavlu V, Yilmaz E. A statistical method for system evaluation using incomplete judgments//Proceedings of the 29th Annual International ACM SIGIR Conference on Research and Development in Information Retrieval. 2006: 541-548.

[50] Robertson S. A new interpretation of average precision//Proceedings of the 31st Annual International ACM SIGIR Conference on Research and Development in Information Retrieval. 2008: 689-690.

[51] Pedregosa F, et al. Scikit-learn: machine learning in python. Journal of Machine Learning Research, 2011, 12: 2825-2830.

[52] Rokach L. Ensemble-based classifiers. Artificial Intelligence Review, 2010, 33(1): 1-39.

[53] Newman M E J. Finding community structure in networks using the eigenvectors of matrices. Physical Review E, 2006, 74(3): 036104.

[54] Shchur O, Mumme M, Bojchevski A, Günnemann S. Pitfalls of graph neural network evaluation. arXiv preprint arXiv:1811.05868, 2019.

[55] Sen P, Namata G, Bilgic M, Getoor L, Galligher B, Eliassi-Rad T. Collective classification in network data. AI Magazine, 2008, 29(3): 93-93.

[56] Bojchevski A, Günnemann S. Deep gaussian embedding of graphs: unsupervised inductive learning via ranking. arXiv preprint arXiv:170, 2017.

[57] Shchur O, Mumme M, Bojchevski A, Günnemann S. Pitfalls of graph neural network evaluation. arXiv preprint arXiv:1811.05868, 2019.

[58] Leskovec J, McAuley J. Learning to discover social circles in ego networks//Advances in Neural Information Processing Systems. 2012, 25.

[59] Wishart D S, et al. DrugBank 5.0: a major update to the DrugBank database for 2018. Nucleic Acids Research, 2018, 46(D1): D1074-D1082.

[60] Hamilton W, Ying Z, Leskovec J. Inductive representation learning on large graphs//Advances in Neural Information Processing Systems. 2017, 30.

[61] Velickovic P, Cucurull G, Casanova A, Romero A, Lio P, Bengio Y, et al. Graph attention networks. Stat, 2017, 1050(20): 10-48550.